\documentclass[aps, prd, showkeys, showpacs ]{revtex4}

\usepackage{amsmath,amsfonts,amscd,amssymb,epsf}
\usepackage[small]{caption}

\newcommand{\pa}{\partial}

\newcommand{\vep}{\varepsilon}

\begin{document}

 \title{ Casimir effect of massive vector fields}
  \author{L.P. Teo}
 \affiliation{Department of Applied Mathematics, Faculty of Engineering, University of Nottingham Malaysia Campus, Jalan Broga, 43500, Semenyih, Selangor Darul Ehsan, Malysia. }\email{LeePeng.Teo@nottingham.edu.my}

\begin{abstract}
We study the Casimir effect due to a massive vector field in a system of two parallel plates made of real materials, in an arbitrary magnetodielectric background. The plane waves satisfying the Proca equations are classified into transverse modes and longitudinal modes which have different dispersion relations. Transverse modes are further divided into type I and type II corresponding to TE and TM modes in the massless case.  For general magnetodielectric media, we argue that the correct boundary conditions are the continuities of $\mathbf{H}_{\parallel}, \phi, \mathbf{A}$ and   $\pa_xA_x$, where $x$ is the direction normal to the plates. Whereas there are type I transverse modes that satisfy all the boundary conditions, it is impossible to find type II transverse modes or longitudinal modes that satisfy all the boundary conditions. To circumvent this problem, type II transverse modes and longitudinal modes have to be considered together. We call the contribution to the Casimir energy from type I transverse modes as TE contribution, and the contribution from the superposition of type II transverse modes and longitudinal modes as TM contribution. Their massless limits give respectively the TE and TM contributions to the Casimir energy of a massless vector field. The limit  where the plates become perfectly conducting is discussed in detail. For the special case where the background has unity refractive index, it is shown that the TM contribution to the Casimir energy  can be written as a sum of contributions from two different types of modes, corresponding to type 2 discrete modes and type 3 continuum modes discussed by Barton and Dombey \cite{1}. For general background, this splitting does not work. The limit  where both plates become infinitely permeable and the limit where one plate becomes perfectly conducting and one plate becomes infinitely permeable are also investigated.
\end{abstract}
\maketitle

\section{Introduction}

In 1948, Casimir \cite{10} predicted the existence of an attractive force between two perfectly conducting plates as a manifestation of the zero-point vacuum fluctuations of quantum fields. Although experimental verification of this prediction has little progress before 1997, Casimir effect has attracted more and more attention since 1970s \cite{17}. It was gradually realized that this effect is closely related to other areas of physics such as quantum field theory, cosmology and condensed matter physics. More recently, the advances in the Casimir measurements and nanotechnology have stimulated more interest in Casimir effect.

Historically, the Casimir effect was investigated for electromagnetic field (massless vector field).
In 1956, Lifshitz \cite{12} derived a formula to express the  Casimir force acting between two dielectric semi-infinite slabs in terms of the permittivity of the slabs \cite{12, 14, 15, 16, 17, 18, 19}. Refinements and generalizations of the Lifshitz formula where the semi-infinite dielectric slabs are replaced by dielectric or magnetodielectric plates of finite thicknesses   were derived in \cite{20,21,22,23,24,25,26,27}. In the limits the  permittivity goes to infinity, Lifshitz formula reproduces the result of Casimir for two perfectly conducting plates.

Since 1970s, Casimir effect was studied for other quantum fields such as scalar fields and spinor fields, for both massless and massive sectors. In contrast, there are very few works that studied Casimir effect of massive vector fields.
One of the incentives to study Casimir effect of massive fields comes from the prevalence of spacetime with extra dimensions which were
proposed to solve some fundamental problems in physics. To study  a quantum field in a spacetime with extra dimensions, one can use Kaluza-Klein decomposition to decompose the quantum field to  an infinite tower of massive fields in $4D$.  This approach has been intrinsically used  in a number of works to study the Casimir effect of scalar field or spinor field in spacetime with extra dimensions such as Kaluza-Klein spacetime and Randall-Sundrum spacetime.  For electromagnetic field, the Casimir effect on a pair of perfectly conducting  plates in the Kaluza-Klein spacetime $M^4\times S^1$ was studied in \cite{11} using the results of Barton and Dombey \cite{1} on Casimir effect of massive photons.

The pioneering work of Barton and Dombey \cite{1,13} considered the Casimir effect on a pair of perfectly conducting plates due to a massive vector field. Unlike the scalar and spinor case, the extension from massless vector field to massive vector field is in fact much more complicated. The classical Maxwell's equations have to be replaced by Proca equations \cite{9} and the gauge degree of freedom is lost. Therefore, a massless vector field has two polarizations but a massive vector field has three.  For a pair of infinitely large perfectly conducting plates, the contribution to the Casimir energy from the two polarizations are the same. However, the contributions to the Casimir energy from the three polarizations are not the same. Another complication in generalizing Casimir effect of massless vector field to massive vector field is that besides the conventional boundary conditions which require the continuities of  $\mathbf{D}_{\perp}, \mathbf{E}_{\parallel}, \mathbf{B}_{\perp}$ and $\mathbf{H}_{\parallel}$ across interfaces, one also need to impose the continuities of the scalar potential $\phi$ and the vector potential $\mathbf{A}$ \cite{5}. For massless vector field where we usually impose the gauge conditions $\phi=0$ and $\nabla\cdot\mathbf{A}=0$, $\mathbf{A}_{\perp}$ is not continuous for the family of TM modes. This gives an obstacle to generalizing the results from the massless case to the massive case. In \cite{1}, Barton and Dombey showed that the eigenmodes of a massive vector field can be divided into two families of discrete modes and one family of continuum modes. In the massless limit, the contribution to the Casimir energy from the discrete modes yields the result of Casimir \cite{10}. The contribution to the Casimir energy from the continuum modes has zero massless limit. One thing that worth remarked is that one family of the discrete modes corresponds to the TE modes in the massless case, but the other family of discrete modes is not a direct generalization of the TM modes in the massless case.

As a first step to investigate the Casimir effect on a pair of plates made of real materials in a spacetime with extra dimensions, we generalize the work of Barton and Dombey \cite{1} to a pair of plates made of real materials in some magnetodielectric background medium. This can also be considered as generalizing the Lifshitz formula for massless vector field to a formula for massive vector field. As explained above, the generalizing of the Casimir effect from massless vector field to massive vector field is nontrivial even for a pair of perfectly conducting plates. Therefore, one should expect that the result will even be more complicated when one considers real materials.
As was discussed in \cite{1}, for dielectric plates, the plane waves of a massive vector field can be divided into two families of transverse waves and one family of longitudinal waves. However, one of the families of discrete modes for perfectly conducting plates is in fact not transverse, and the continuum modes for perfectly conducting plates is in fact not longitudinal. Therefore, it is not obvious how one should generalize the work of Barton and Dombey \cite{1}. In fact, one can show that for plates made of real materials in a magnetodielectric background, only one family of the transverse modes can satisfy all the boundary conditions.  As a result, to look for other eigenmodes of the massive vector field, we are forced to consider the superposition of the other family of transverse modes and the family of longitudinal modes.

In \cite{12}, Lifshitz modeled two dielectric plates as two semi-infinite slabs which is a reasonable model when the skin depths of the materials of the plates are much smaller than the thicknesses of the plates. The configuration of two semi-infinite slabs separated by a medium is   a three-layer model. However, since we want to consider plates made of any materials, we find it necessary not to make any assumption about the skin depths. Therefore, we would consider two plates with finite thicknesses which can be modeled by a five-layer configuration.

The layout of this article is as follows. In Section \ref{s2}, we review the Proca equations for massive vector fields. In Section \ref{s3}, we divide the plane waves solutions to the Proca equations into type I and type II transverse waves, and longitudinal waves. In Section \ref{s4}, we discuss the boundary conditions that should be satisfied by the potentials and fields on the interface of two plane parallel media. In Section \ref{s5}, we derive the Casimir energies contributed by transverse modes of type I, and by the superposition of transverse modes of type II and longitudinal modes. We show that their massless limits are the TE and TM contributions to the Casimir energy due to a massless vector field. In Section \ref{s6}, we discuss the limiting case where the plates become perfectly conducting. In Section \ref{s7}, we consider the limit when the plates become infinitely permeable. In Section \ref{s8}, we study the case where one plate is perfectly conducting and one plate is infinitely permeable.

\section{Proca equations   of   massive vector field  in a magnetodielectric medium}\label{s2}
A massive vector field is represented by a  four-vector $\displaystyle\left\langle \frac{\phi}{c}, A_x, A_y, A_z\right\rangle$, where $\phi$ is the scalar potential and $\mathbf{A}=\langle A_x, A_y, A_z\rangle$ is the vector potential. Define the electric field $\mathbf{E}$ and the magnetic field $\mathbf{B}$ by \begin{align}\label{eq1_5}\mathbf{E}=-\frac{\pa\mathbf{A}}{\pa t}-\nabla\phi,\hspace{1cm}\mathbf{B}=\nabla\times \mathbf{A}.\end{align} Consider a magnetodielectric medium with    permittivity $\varepsilon$ and permeability $\mu$. Let $\rho_f$ and $\mathbf{J}_f$ be the free charge and free current in the medium.  Assume the usual linear relations $\mathbf{D}=\varepsilon \mathbf{E}$ and $\mathbf{B}=\mu\mathbf{H}$. Then the Proca equations for the electromagnetic waves propagating in this medium are \cite{1}:
\begin{align}
&\nabla\cdot\mathbf{B}=0,\label{eq1_1}\\
&\nabla\times\mathbf{E}+\frac{\pa\mathbf{B}}{\pa t}=0,\label{eq1_2}\\
&\nabla\cdot\mathbf{D}+\frac{m^2 }{\mu\hbar^2}\phi=  \rho_f, \label{eq1_3}\\
&\nabla\times \mathbf{H}-\frac{\pa\mathbf{D}}{\pa t}+ \frac{m^2c^2}{ \mu\hbar^2}\mathbf{A}=\mathbf{J}_f.\label{eq1_4}
\end{align}The first two are the well-known Maxwell's equations which are automatically satisfied because of \eqref{eq1_5}.
   The conservation of free charges implies the continuity equation:
\begin{align}\label{eq6_24_1}
\frac{\pa\rho_f}{\pa t}+\nabla\cdot\mathbf{J}_f=0.
\end{align}Differentiate \eqref{eq1_3} with respect to $t$ and apply the divergence operator to \eqref{eq1_4}, \eqref{eq6_24_1} implies that
 the Lorentz condition
\begin{align}\label{eq1_6}
\frac{1}{c^2}\frac{\pa\phi}{\pa t}+ \nabla\cdot\mathbf{A}=0
\end{align}has to be satisfied. From \eqref{eq1_5} and \eqref{eq1_6}, one can then derive from \eqref{eq1_3} and \eqref{eq1_4}   the following two equations for $\phi$ and $\mathbf{A}$:
\begin{align}
\left(\frac{1}{c^2}\frac{\pa^2}{\pa t^2}-\nabla^2+\frac{m^2 }{\varepsilon\mu\hbar^2}\right)\phi=&\frac{\rho_f}{\varepsilon},
\label{eq1_7}\\
\left(\varepsilon\mu\frac{\pa^2}{\pa t^2}-\nabla^2+ \frac{m^2c^2}{ \hbar^2}\right)\mathbf{A}-\left( \varepsilon\mu c^2 -1\right)\nabla(\nabla\cdot\mathbf{A})=&\mu\mathbf{J}_f.\label{eq1_8}
\end{align}
\eqref{eq1_6}, \eqref{eq1_7} and \eqref{eq1_8} are the equivalences of the Proca equations for the potentials $\phi$ and $\mathbf{A}$.

For a massless vector field, the primary quantities in the Maxwell's equations are the electric and magnetic fields. There is a gauge degree of freedom given by
$$\mathbf{A}\mapsto \mathbf{A}+\nabla \psi,\quad \phi\mapsto \phi - \frac{\pa\psi}{\pa t}$$for an arbitrary function $\psi$. The Lorentz gauge \eqref{eq1_6} is one of the gauge conditions that can be used to fix the gauge. This gauge degree of freedom is lost in the massive case. In fact, for massive vector field, the primary quantities are the potentials $\phi$ and $\mathbf{A}$, and the electric and magnetic fields are derived quantities. The Lorentz condition \eqref{eq1_6} is a necessary condition followed from the conservation of charges.

\section{Plane waves and dispersionless relations}\label{s3}
Consider an unbounded magnetodielectric medium with permittivity $\vep$ and permeability $\mu$ in the absence of free chargers and currents, i.e., $\rho_f=0$ and $\mathbf{J}_f=0$. As discussed in \cite{1},   the monochromatic plane wave solutions of the Proca equations can be divided into transverse waves with $\nabla\cdot\mathbf{A}=0$ and longitudinal waves with $\nabla\times\mathbf{A}=\mathbf{0}$. For the transverse waves, it follows from Lorentz condition \eqref{eq1_6}  that $\phi=0$. The equation \eqref{eq1_8} for the vector potential $\mathbf{A}$ becomes \begin{equation}\label{eq7_20_1}\left(\varepsilon\mu\frac{\pa^2}{\pa t^2}-\nabla^2+ \frac{m^2c^2}{ \hbar^2}\right)\mathbf{A}=\mathbf{0}.\end{equation} Recall that for a massless vector field, the  vector potential $\mathbf{A}$ satisfies the equation \eqref{eq7_20_1} with $m=0$, and the conditions $\phi=0$ and $\nabla \cdot \mathbf{A}=0$, called Coulomb gauge or radiation gauge, are usually imposed to remove the gauge degree of freedom. Therefore the transverse waves are in one-to-one correspondence with the massless electromagnetic waves. Since massless electromagnetic waves are usually divided into TE and TM polarizations, we do the same for the massive case, but call them type I and type II transverse waves.

In the following, let $\mathbf{k}_{\perp}=(k_2, k_3)$, $k_{\perp}=\sqrt{k_2^2+k_3^2}$ and define $f_{\mathbf{k}_{\perp},\omega}(y,z,t)=e^{ik_2y+ik_3z-i\omega t}$.

\vspace{0.3cm}\noindent
\textbf{A. Type I transverse  waves.} The type I transverse   waves are the equivalence of the TE waves in the massless case:
\begin{align*}
\left\{\begin{aligned}A_x=&0\\
A_y=& -k_3 e^{ip_Tx}  f_{\mathbf{k}_{\perp},\omega}(y,z,t)\\
A_z=& k_2 e^{ip_Tx} f_{\mathbf{k}_{\perp},\omega}(y,z,t)\\
\phi=&0 \end{aligned}\right..
\end{align*} In this case, $E_x=0$.
The   dispersion relation is:
$$-\epsilon\mu\omega^2+p_T^2+k_{\perp}^2+\frac{m^2c^2}{\hbar^2}=0.$$

\vspace{0.3cm}\noindent
\textbf{B. Type II transverse  waves.} The type II transverse   waves are the equivalence of the TM waves in the massless case:
\begin{align*}
\left\{\begin{aligned}A_x=&\frac{k_{\perp}^2}{p_T} e^{ip_Tx}f_{\mathbf{k}_{\perp},\omega}(y,z,t)\\
A_y=& - k_2  e^{ip_Tx}  f_{\mathbf{k}_{\perp},\omega}(y,z,t)\\
A_z=& - k_3   e^{ip_Tx} f_{\mathbf{k}_{\perp},\omega}(y,z,t)\\
\phi=&0\end{aligned}\right.
\end{align*}  In this case, $B_x=0$. The   dispersion relation is:
$$-\varepsilon\mu\omega^2+p_T^2+k_{\perp}^2+\frac{m^2c^2}{\hbar^2}=0.$$

\vspace{0.3cm}\noindent
\textbf{C. Longitudinal waves.} The longitudinal waves are
\begin{align*}
\left\{\begin{aligned}A_x=&  p_L e^{ip_Lx} f_{\mathbf{k}_{\perp},\omega}(y,z,t)\\
A_y=&  k_2 e^{ip_Lx} f_{\mathbf{k}_{\perp},\omega}(y,z,t)\\
A_z=&  k_3 e^{ip_Lx} f_{\mathbf{k}_{\perp},\omega}(y,z,t)\\
\phi=&\frac{c^2  }{\omega}(p_L^2+k_{\perp}^2) e^{ip_Lx} f_{\mathbf{k}_{\perp},\omega}(y,z,t)\end{aligned}\right.,
\end{align*}
The   dispersion relation is:
\begin{align*}&-\frac{\omega^2}{c^2}+p_L^2+k_{\perp}^2+\frac{m^2 }{\varepsilon\mu\hbar^2}= 0.\end{align*}For longitudinal waves, $\nabla\times\mathbf{A}=\mathbf{0}$ implies that the magnetic field vanishes identically, i.e., $\mathbf{B}=0$. On the other hand, the electric field is given by
\begin{align*}
\left\{\begin{aligned}E_x=&\frac{ip_L c^2}{\omega }\left(\frac{\omega^2}{c^2}-p_{L}^2-k_{\perp}^2\right) e^{ip_Lx} f_{\mathbf{k}_{\perp},\omega}(y,z,t)=\frac{ip_L m^2c^2}{\vep\mu\omega\hbar^2 }e^{ip_Lx} f_{\mathbf{k}_{\perp},\omega}(y,z,t)\\
E_y=&\frac{ik_2 c^2}{\omega }\left( \frac{\omega^2}{c^2}-p_{L}^2-k_{\perp}^2\right) e^{ip_Lx} f_{\mathbf{k}_{\perp},\omega}(y,z,t)=\frac{ik_2 m^2c^2}{\vep\mu\omega\hbar^2 }e^{ip_Lx} f_{\mathbf{k}_{\perp},\omega}(y,z,t)\\
E_z=&\frac{ik_3 c^2}{\omega }\left(\frac{\omega^2}{c^2}-p_{L}^2-k_{\perp}^2\right) e^{ip_Lx} f_{\mathbf{k}_{\perp},\omega}(y,z,t)=\frac{ik_3m^2 c^2}{\vep\mu\omega\hbar^2 }e^{ip_Lx} f_{\mathbf{k}_{\perp},\omega}(y,z,t)
\end{aligned}\right..
\end{align*}Notice that the electric field also vanishes in the massless $m\rightarrow 0$ limit. It should be observed that the longitudinal waves and transverse waves satisfy different dispersion  relations, unless $\vep\mu=1/c^2$, i.e., the waves traveled at the speed of light in the medium.

\section{Boundary conditions}\label{s4}
In this section, we consider  the boundary conditions that must be satisfied by  the potentials $\mathbf{A}$ and $\phi$ and the   fields $\mathbf{E}$ and $\mathbf{B}$ on the interface of two plane parallel  media in the absence of free charges and currents. Let $x$ be the direction normal to the boundary of the two media. As in the massless case,
\eqref{eq1_1} and \eqref{eq1_2} imply that $\mathbf{B}_{\perp}=B_x$ and $ \mathbf{E}_{\parallel}=(E_y, E_z)$ are continuous.
  Taking the derivative of \eqref{eq1_3} with respect to $t$ gives
\begin{align*}
&\nabla\cdot\left(\frac{\pa \mathbf{D}}{\pa t}\right)+\frac{m^2c^2}{\mu \hbar^2}\frac{1}{c^2}\frac{\pa\phi}{\pa t}=0.\end{align*}The   Lorentz condition \eqref{eq1_6} then imply that
\begin{align*}
 \nabla\cdot\left(\frac{\pa \mathbf{D}}{\pa t} -\frac{m^2c^2}{\mu\hbar^2}\mathbf{A} \right)=0.
\end{align*}Therefore, we find that
\begin{align*}
\left(\frac{\pa \mathbf{D}}{\pa t}  -\frac{m^2c^2}{\mu\hbar^2}\mathbf{A}\right)_{\perp}=\frac{\pa\left(\vep E_x\right)}{\pa t} -\frac{m^2c^2}{\mu\hbar^2}A_x
\end{align*}must be continuous.
\eqref{eq1_4} then implies that $\displaystyle\mathbf{H}_{\parallel}=\left(\frac{B_y}{\mu},\frac{B_z}{\mu}\right)$ has to be continuous. Notice that  the boundary condition that one obtains from \eqref{eq1_3} is the continuity of
$$\left(\frac{\pa \mathbf{D}}{\pa t}  -\frac{m^2c^2}{\mu\hbar^2}\mathbf{A}\right)_{\perp},$$ but not the continuity of $\mathbf{D}_{\perp}$ as in the massless case.

As discussed in \cite{1}, the boundary conditions above are not enough for massive vector fields. It was pointed out by Kroll \cite{5} that the potentials $\phi$ and $\mathbf{A}$ also have to be continuous. Therefore, the complete set of boundary conditions is the continuities of $\displaystyle B_x, E_y, E_z, \frac{\pa\left(\vep E_x\right)}{\pa t} -\frac{m^2c^2}{\mu\hbar^2}A_x, \frac{B_y}{\mu},\frac{B_z}{\mu}, A_x, A_y, A_z, \phi$.
However, not all these conditions are independent. For example,  the continuities of $\phi$ and $\mathbf{A}_{\parallel}=(A_y, A_z)$ imply the continuity of$$\mathbf{E}_{\parallel}=-\nabla_{\parallel} \phi-\frac{\pa\mathbf{A}_{\parallel}}{\pa t}.$$The continuities of $\displaystyle \frac{B_y}{\mu},\frac{B_z}{\mu}$ imply the continuity of  $\displaystyle \frac{\pa\left(\vep E_x\right)}{\pa t} -\frac{m^2c^2}{\mu\hbar^2}A_x$ since \eqref{eq1_4} gives
$$\frac{\pa\left(\vep E_x\right)}{\pa t} -\frac{m^2c^2}{\mu\hbar^2}A_x=\frac{1}{\mu}\left(\frac{\pa B_z}{\pa y}-\frac{\pa B_y}{\pa z}\right).$$Similarly, \eqref{eq1_2} implies that if $\mathbf{E}_{\parallel}=(E_y,E_z)$ is continuous, then so is $B_x$. On the other hand, the continuities of $\mathbf{A}_{\parallel}=(A_y, A_z)$ and the Lorentz condition
 $$0=\frac{1}{c^2}\frac{\pa\phi}{\pa t}+\nabla\cdot \mathbf{A}=\frac{1}{c^2}\frac{\pa\phi}{\pa t} + \frac{\pa A_x}{\pa x}+\nabla_{\parallel}\cdot\mathbf{A}_{\parallel}$$ imply that $\phi$ is continuous if and only if $\displaystyle \frac{\pa A_x}{\pa x}$ is continuous.

In conclusion, for the boundary conditions, it is sufficient to  impose the continuities of $\displaystyle  A_x, A_y, A_z, \frac{B_y}{\mu},\frac{B_z}{\mu}$ and the continuity of either $\phi$ \emph{or} $\displaystyle\frac{\pa A_x}{\pa x}$.

\section{Casimir energy of parallel magnetodielectric plates inside magnetodielectric medium}\label{s5}
\begin{figure}[h]
\epsfxsize=0.4\linewidth \epsffile{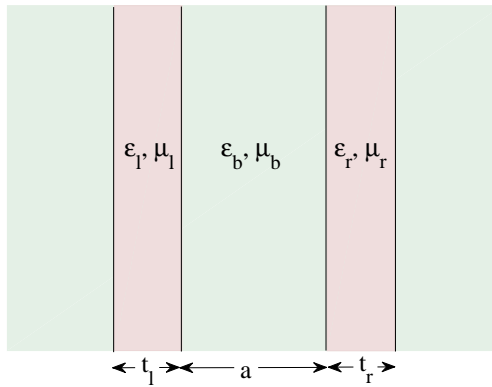} \caption{\label{f2}Two parallel magnetodielectric plates inside a magnetodielectric medium. }\end{figure}
\begin{figure}[h]
\epsfxsize=0.4\linewidth \epsffile{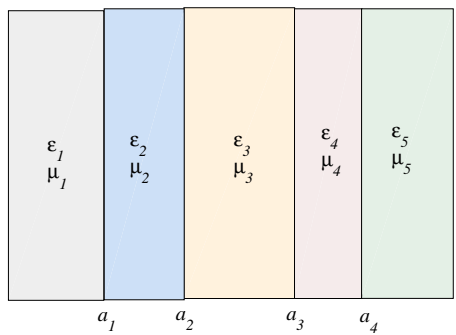} \caption{\label{f1}A five-layer model }\end{figure}

In this article, we are interested in computing the Casimir energy of two parallel magnetodielectric plates inside a magnetodielectric medium (see FIG. \ref{f2}). We assume that the cross section of the plates are infinite. For convenience, we first consider a five-layer model consists of five plane-parallel layers of magnetodielectric media as shown in FIG. \ref{f1}.
At the end, we only consider the case where $\vep_1=\vep_3=\vep_5$ and $\mu_1=\mu_3=\mu_5$. The interfaces of the media are located at $x=a_1, a_2, a_3$ and $a_4$, with $a_1<a_2<a_2<a_4$. We  assume that there are two artificial boundaries at $x=a_0<a_1$ and $x=a_5>a_4$, and let $d_i=a_i-a_{i-1}$, $1\leq i\leq 5$. At the end, we are going to let $d_1, d_5\rightarrow \infty$. We also set $t_l=d_2$ and $t_r=d_4$ to be the thicknesses of the left and right plates that we are interested in, and set $a=d_3$ to be the separation between the plates.

For the Casimir energy of massless electromagnetic field in the five layer model, one can consider the contribution from TE modes and TM modes separately. However, it turns out that for the massive case, one \emph{cannot} consider the contributions from the three types of waves  separately. In fact,   the contribution from transverse modes of type I can be considered independently, but  \emph{the contributions from the transverse modes of type II and the longitudinal modes have to be considered} \emph{together}. It is impossible to find type II transverse modes or longitudinal modes that   satisfy all the boundary conditions.

\subsection{Contribution from transverse modes of type I (TE modes)}
For transverse modes of type I, assume that
\begin{align*}
\left\{\begin{aligned}A_x=&0\\
A_y=& -k_3(\mathfrak{A}_ie^{ip_{T,i}x}+\mathfrak{B}_ie^{-ip_{T,i}x}) f_{\mathbf{k}_{\perp},\omega}(y,z,t)\\
A_z=& k_2(\mathfrak{A}_ie^{ip_{T,i}x}+\mathfrak{B}_ie^{-ip_{T,i}x}) f_{\mathbf{k}_{\perp},\omega}(y,z,t)\\
\phi=&0 \end{aligned}\right.\quad \text{for} \;a_{i-1}\leq x\leq a_i.
\end{align*} The dispersion relation implies that
$$p_{T,i}^2=\vep_i\mu_i\omega^2-k_{\perp}^2-\frac{m^2c^2}{\hbar^2}.$$
The electric field $\mathbf{E}$ and the magnetic field $\mathbf{B}$ are given respectively by
\begin{align*}
\left\{\begin{aligned}E_x=&0\\
E_y=&-i\omega k_3(\mathfrak{A}_ie^{ip_{T,i}x}+\mathfrak{B}_ie^{-ip_{T,i}x}) f_{\mathbf{k}_{\perp},\omega}(y,z,t)\\
E_z=&i\omega k_2(\mathfrak{A}_ie^{ip_{T,i}x}+\mathfrak{B}_ie^{-ip_{T,i}x}) f_{\mathbf{k}_{\perp},\omega}(y,z,t)
\end{aligned}\right.,
\hspace{1cm}
\left\{\begin{aligned}B_x=&ik_{\perp}^2(\mathfrak{A}_ie^{ip_{T,i}x}+\mathfrak{B}_ie^{-ip_{T,i}x}) f_{\mathbf{k}_{\perp},\omega}(y,z,t)\\
B_y=& -ip_{T,i}k_2(\mathfrak{A}_ie^{ip_{T,i}x}-\mathfrak{B}_ie^{-ip_{T,i}x}) f_{\mathbf{k}_{\perp},\omega}(y,z,t)\\
B_z=&-ip_{T,i}k_3(\mathfrak{A}_ie^{ip_{T,i}x}-\mathfrak{B}_ie^{-ip_{T,i}x}) f_{\mathbf{k}_{\perp},\omega}(y,z,t)
\end{aligned}\right.
\end{align*}for $a_{i-1}<x<a_i$. In this case, $\displaystyle \phi,  A_x$ are automatically continuous.
The continuities of $  A_y, A_z$    imply that
  for $i=1,2,3,4$,
\begin{equation*}
\begin{split}
&\mathfrak{A}_ie^{ip_{T,i}a_i}+\mathfrak{B}_ie^{-ip_{T,i}a_i} =\mathfrak{A}_{i+1}e^{ip_{T,i+1}a_i}+\mathfrak{B}_{i+1}e^{-ip_{T,i+1}a_i}.
\end{split}
\end{equation*}The continuities of $\displaystyle \frac{B_y}{\mu}, \frac{B_z}{\mu}$ imply that
  for $i=1,2,3,4$,
\begin{equation*}
\begin{split}
&\frac{p_{T,i}}{\mu_i}\left[\mathfrak{A}_ie^{ip_{T,i}a_i}-\mathfrak{B}_ie^{-ip_{T,i}a_i}\right] =
\frac{p_{T,i+1}}{\mu_{i+1}}\left[\mathfrak{A}_{i+1}e^{ip_{T,i+1}a_i}-\mathfrak{B}_{i+1}e^{-ip_{T,i+1}a_i}\right].
\end{split}
\end{equation*}Next, we have to impose some boundary conditions on the artificial boundaries $x=a_0$ and $x=a_5$. As in the massless case, we impose the conditions $E_y=E_z=0$ and $B_x=0$, which gives
$$\mathfrak{A}_1e^{ip_{T,1}a_0}+\mathfrak{B}_1e^{-ip_{T,1}a_0}=0,\quad\mathfrak{A}_5e^{ip_{T,5}a_5}+\mathfrak{B}_5e^{-ip_{T,5}a_5}=0.$$
From here,  one can use the method we developed in our previous work \cite{2}. After a renormalization by the criterion that the Casimir energy should vanish in the limit the plate separation $d_3$ is infinite, and passing to the limits $d_1,d_5\rightarrow \infty$, one finds that   the TE contribution to the   Casimir energy is given by
\begin{equation}\label{eq7_22_2}
\begin{split}
E_{\text{Cas}}^{\text{TE}} =& \frac{\hbar}{4\pi^2}\int_0^{\infty}\int_0^{\infty} \ln\Biggl\{
1-\mathfrak{D}^{\text{TE}}(\xi,k_\perp)e^{-2q_{T,b}(\xi,k_\perp)a}\Biggr\} k_{\perp}dk_{\perp}d\xi,
\end{split}
\end{equation}where
\begin{equation}\label{eq7_22_1}\mathfrak{D}^{\text{TE}}(\xi,k_\perp)=\frac{\Delta_l^{\text{I}}\Delta_r^{\text{I}}\left(1-e^{-q_{T,l}t_l} \right)
\left(1-e^{-q_{T,r}t_r}\right)}{\left(1-\left[\Delta_l^{\text{I}}\right]^2e^{-q_{T,l}t_l} \right)
\left(1-\left[\Delta_{r}^{\text{I}}\right]^2e^{-q_{T,r}t_r}\right)}.\end{equation}
Here  the variables have been renamed by $\vep_1=\vep_3=\vep_5=\vep_b$, $\mu_1=\mu_3=\mu_5=\mu_b$, $\vep_l=\vep_2, \vep_r=\vep_4, \mu_l=\mu_2, \mu_r=\mu_4$, $t_l=d_2, t_r=d_4, a=d_3$; and
\begin{equation}\label{eq9_6_1}
\begin{split}
 q_{T,j}(\xi,k_\perp)=& \sqrt{\vep_j(i\xi)\mu_j(i\xi)\xi^2+k_{\perp}^2+\frac{m^2c^2}{\hbar^2}},\quad j=b,l,r;\\
\Delta_j^{\text{I}}(\xi,k_\perp)=& \frac{q_{T,j}(\xi,k_{\perp})\mu_b(i\xi)-q_{T,b}(\xi,k_{\perp})\mu_j(i\xi)}{q_{T,j}(\xi, k_{\perp})\mu_b(i\xi)+q_{T,b}(\xi, k_{\perp})\mu_j(i\xi)},\quad j=l,r.
\end{split}
\end{equation}
The TE contribution to the Casimir force is given by
\begin{equation*}
F_{\text{Cas}}^{\text{TE} } =-\frac{\pa E_{\text{Cas}}^{\text{TE} }}{\pa a}=-\frac{\hbar}{ 2\pi^2 }\int_0^{\infty}\int_0^{\infty} q_{T,b}(\xi,k_\perp)\left(
 \left[\mathfrak{D}^{\text{TE}}(\xi,k_\perp)\right]^{-1}e^{2q_{T,b}(\xi,k_\perp)a}-1\right)^{-1} k_{\perp}dk_{\perp}d\xi.
\end{equation*}
Obviously, in the massless $m\rightarrow 0$ limit, one obtains the result for the massless case, where the mass $m$ in $q_{T,j}(\xi,k_{\perp})$ in \eqref{eq9_6_1} is set to zero.

\subsection{Contributions from superposition of transverse modes of type II and longitudinal modes (TM modes)}
For the superposition of transverse modes of type II and longitudinal modes, we have for $a_{i-1}<x<a_i$,
\begin{equation}\label{eq7_22_9}\begin{split}
\left\{\begin{aligned}A_x=&\left[\frac{k_{\perp}^2}{p_{T,i}}(\mathfrak{C}_ie^{ip_{T,i}x}-\mathfrak{D}_ie^{-ip_{T,i}x})+ p_{L,i}(\mathfrak{E}_ie^{ip_{L,i}x}-\mathfrak{F}_ie^{-ip_{L,i}x})\right] f_{\mathbf{k}_{\perp},\omega}(y,z,t)\\
A_y=&\left[  - k_2  (\mathfrak{C}_ie^{ip_{T,i}x}+\mathfrak{D}_ie^{-ip_{T,i}x})+ k_2(\mathfrak{E}_ie^{ip_{L,i}x}+\mathfrak{F}_ie^{-ip_{L,i}x})\right]f_{\mathbf{k}_{\perp},\omega}(y,z,t)\\
A_z=&\left[ - k_3 (\mathfrak{C}_ie^{ip_{T,i}x}+\mathfrak{D}_ie^{-ip_{T,i}x})
+  k_3(\mathfrak{E}_ie^{ip_{L,i}x}+\mathfrak{F}_ie^{-ip_{L,i}x})\right]f_{\mathbf{k}_{\perp},\omega}(y,z,t)\\
\phi=&\frac{c^2  }{\omega}(p_{L,i}^2+k_{\perp}^2)(\mathfrak{E}_ie^{ip_{L,i}x}+\mathfrak{F}_ie^{-ip_{L,i}x})f_{\mathbf{k}_{\perp},\omega}(y,z,t) \end{aligned}\right..
\end{split}\end{equation} The electric field $\mathbf{E}$ and the magnetic field $\mathbf{B}$ are given respectively by
\begin{align}\label{eq7_22_11}
\left\{\begin{aligned}E_x=&i\left[\frac{\omega k_{\perp}^2}{p_{T,i}}(\mathfrak{C}_ie^{ip_{T,i}x}-\mathfrak{D}_ie^{-ip_{T,i}x})+\frac{p_{L,i} m^2c^2}{\vep_i\mu_i\omega \hbar^2 } (\mathfrak{E}_ie^{ip_{L,i}x}-\mathfrak{F}_ie^{-ip_{L,i}x})\right]f_{\mathbf{k}_{\perp},\omega}(y,z,t)\\
E_y=&i\left[  -\omega  k_2 (\mathfrak{C}e^{ip_{T,i}x}+\mathfrak{D}e^{-ip_{T,i}x}) +\frac{k_2 m^2c^2}{\vep_i\mu_i\omega \hbar^2 } (\mathfrak{E}_ie^{ip_{L,i}x}+\mathfrak{F}_ie^{-ip_{L,i}x})\right] f_{\mathbf{k}_{\perp},\omega}(y,z,t)\\
E_z=&i\left[  - \omega k_3 (\mathfrak{C}e^{ip_{T,i}x}+\mathfrak{D}e^{-ip_{T,i}x})
+\frac{k_3 m^2c^2}{\vep_i\mu_i\omega \hbar^2 } (\mathfrak{E}_ie^{ip_{L,i}x}+\mathfrak{F}_ie^{-ip_{L,i}x})\right] f_{\mathbf{k}_{\perp},\omega}(y,z,t)
\end{aligned}\right.
\end{align}
\begin{align}\label{eq7_22_12}
\left\{\begin{aligned}B_x=&0\\
B_y=& \frac{i  k_3}{p_{T,i}} \left(p_{T,i}^2+k_{\perp}^2\right)(\mathfrak{C}_ie^{ip_{T,i}x}-\mathfrak{D}_ie^{-ip_{T,i}x}) f_{\mathbf{k}_{\perp},\omega}(y,z,t)\\
B_z=& -\frac{i  k_2}{p_{T,i}} \left(p_{T,i}^2+k_{\perp}^2\right)(\mathfrak{C}_ie^{ip_{T,i}x}-\mathfrak{D}_ie^{-ip_{T,i}x})  f_{\mathbf{k}_{\perp},\omega}(y,z,t)
\end{aligned}\right..
\end{align} Since $B_x=0$, we can call these modes   TM modes.  The continuities of   $\displaystyle    \frac{B_y}{\mu}, \frac{B_z}{\mu},\phi, A_x, A_y, A_z  $  imply that
for $i=1,2,3,4$,
\begin{equation}\label{eq7_22_10}\begin{split}
& \frac{1}{\mu_ip_{T,i}}\left(p_{T,i}^2+k_{\perp}^2\right)(\mathfrak{C}_ie^{ip_{T,i}a_i}-\mathfrak{D}_ie^{-ip_{T,i}a_i})
= \frac{1}{\mu_{i+1}p_{T,i+1}}\left(p_{T,i+1}^2+k_{\perp}^2\right)(\mathfrak{C}_{i+1}e^{ip_{T,i+1}a_i}-\mathfrak{D}_{i+1}e^{-ip_{T,i+1}a_i})\\
& \left(p_{L,i}^2+k_{\perp}^2\right)(\mathfrak{E}_ie^{ip_{L,i}a_i}+\mathfrak{F}_ie^{-ip_{L,i}a_i})
= \left(p_{L,i+1}^2+k_{\perp}^2\right)(\mathfrak{E}_{i+1}e^{ip_{L,i+1}a_i}+\mathfrak{F}_{i+1}e^{-ip_{L,i+1}a_i})\\
& \frac{k_{\perp}^2}{p_{T,i}}(\mathfrak{C}_ie^{ip_{T,i}a_i}-\mathfrak{D}_ie^{-ip_{T,i}a_i})+ p_{L,i}(\mathfrak{E}_ie^{ip_{L,i}a_i}-\mathfrak{F}_ie^{-ip_{L,i}a_i})\\&\hspace{0.5cm}=
\frac{k_{\perp}^2}{p_{T,i+1}}(\mathfrak{C}_{i+1}e^{ip_{T,i+1}a_i}-\mathfrak{D}_{i+1}e^{-ip_{T,i+1}a_i})+ p_{L,i+1}(\mathfrak{E}_{i+1}e^{ip_{L,i+1}a_i}-\mathfrak{F}_{i+1}e^{-ip_{L,i+1}a_i})
\\
& (\mathfrak{C}_ie^{ip_{T,i}a_i}+\mathfrak{D}_ie^{-ip_{T,i}a_i}) - (\mathfrak{E}_ie^{ip_{L,i}a_i}+\mathfrak{F}_ie^{-ip_{L,i}a_i})
\\&\hspace{0.5cm}= (\mathfrak{C}_{i+1}e^{ip_{T,i+1}a_i}+\mathfrak{D}_{i+1}e^{-ip_{T,i+1}a_i}) - (\mathfrak{E}_{i+1}e^{ip_{L,i+1}a_i}+\mathfrak{F}_{i+1}e^{-ip_{L,i+1}a_i}).
\end{split}\end{equation}
For the artificial boundaries at $x=a_0$ and $x=a_5$, one can impose the conditions $A_y=A_z=\phi=0$. These imply that
\begin{align*}
 \mathfrak{C}_1e^{ ip_{T,1}a_0}+\mathfrak{D}_1e^{ -ip_{T,1}a_0} =&0, \quad
 \mathfrak{E}_1e^{ip_{L,1}a_0}+\mathfrak{F}_1e^{-ip_{L,1}a_0} =0,\\
 \mathfrak{C}_5e^{ip_{T,5}a_5}+\mathfrak{D}_5e^{ -ip_{T,5}a_5} =&0,\quad
\mathfrak{E}_5e^{ip_{L,5}a_5}+\mathfrak{F}_5e^{-ip_{L,5}a_5} =0.
\end{align*}
As in the case of TE modes, one can  then show  that the TM contribution to the   Casimir energy is given   by
\begin{equation}\label{eq7_22_5}
\begin{split}
E_{\text{Cas}}^{\text{TM}} =& \frac{\hbar}{4\pi^2}\int_0^{\infty}\int_0^{\infty} \ln\det \frac{Q(\xi, k_{\perp})}{Q_{\infty}(\xi, k_{\perp})}d\xi k_{\perp}dk_{\perp}.
\end{split}
\end{equation}
With the notations
\begin{equation*}
\begin{split}
q_{L,j}(\xi,k_{\perp})=& \sqrt{\frac{\xi^2}{c^2}+k_{\perp}^2+\frac{m^2}{\vep_j(i\xi)\mu_j(i\xi)\hbar^2}}, \quad j=b,l,r;\\ \Delta_{j}^{\text{II}}(\xi,k_{\perp})
=& \frac{\mu_b(i\xi)q_{T,b}(\xi,k_{\perp})[q_{T,j}(\xi,k_{\perp})^2-k_{\perp}^2]-\mu_j(i\xi)q_{T,j}(\xi,k_{\perp})[q_{T,b}(\xi,k_{\perp})^2-k_{\perp}^2]}
{\mu_b(i\xi)q_{T,b}(\xi,k_{\perp})[q_{T,j}(\xi,k_{\perp})^2-k_{\perp}^2]+\mu_j(i\xi)q_{T,j}(\xi,k_{\perp})[q_{T,b}(\xi, k_{\perp})^2-k_{\perp}^2]},\quad j=l,r;\end{split}
\end{equation*}\begin{equation*}
\begin{split}
\Delta_{j}^{\text{III}}(\xi,k_{\perp})=& \frac{q_{L,j}(\xi,k_{\perp})[q_{L,b}(\xi,k_{\perp})^2-k_{\perp}^2]-q_{L,b}(\xi,k_{\perp})[q_{L,j}(\xi,k_{\perp})^2-k_{\perp}^2]}
{q_{L,j}(\xi,k_{\perp})[q_{L,b}(\xi,k_{\perp})^2-k_{\perp}^2]+q_{L,b}(\xi,k_{\perp})[q_{L,j}(\xi,k_{\perp})^2-k_{\perp}^2]},\quad j=l,r;\\
\alpha_{j_1j_2}(\xi,k_{\perp})=& \frac{q_{L,j_1}(\xi,k_{\perp})^2-q_{L,j_2}(\xi,k_{\perp})^2}{q_{L,j_2}(\xi,k_{\perp})^2-k_{\perp}^2},\quad
j_1,j_2=b,l,r;\\
\beta_{j_1j_2}(\xi,k_{\perp})=& -\frac{k_{\perp}^2}{q_{T,j_1}(\xi,k_{\perp})q_{L,j_2}(\xi,k_{\perp})}\left(1-\frac{\mu_{j_2}(i\xi)}{\mu_{j_1}(i\xi)}\frac{[q_{T,j_1}(\xi,k_{\perp})^2
-k_{\perp}^2]}{[q_{T,j_2}(\xi,k_{\perp})^2-k_{\perp}^2]}\right),\quad
j_1,j_2=b,l,r;\\
r_{j_1j_2}^+(\xi,k_{\perp})=& 1+\frac{q_{T,j_2}(\xi,k_{\perp})\mu_{j_2}(i\xi)}{q_{T,j_1}(\xi,k_{\perp})\mu_{j_1}(i\xi)}\frac{[q_{T,j_1} (\xi,k_{\perp})^2-k_{\perp}^2]}
{[q_{T,j_2}(\xi,k_{\perp})^2-k_{\perp}^2]},\quad j_1,j_2=b,l,r;\\
\kappa_{j_1j_2}^+(\xi,k_{\perp})=& \frac{q_{L,j_1}(\xi,k_{\perp})^2-k_{\perp}^2}{q_{L,j_2}(\xi,k_{\perp})^2-k_{\perp}^2}+\frac{q_{L,j_1}(\xi,k_{\perp})}{q_{L,j_2}(\xi,k_{\perp})}, \quad j_1,j_2=b,l,r;
\end{split}
\end{equation*}
   $Q$ is a $4\times 4$ matrix with components given by
\begin{equation*}
\begin{split}
Q_{11}= &\left[-\Delta_{l}^{\text{II}}\left(1- e^{-2q_{T,l}t_l}    \right) e^{q_{T,l}t_l}+\frac{\alpha_{lb}}{r_{lb}^+}\frac{\beta_{bl}}{r_{bl}^+}\left(1-e^{-2q_{L,l}t_l} \right)e^{q_{L,l}t_l}\right] e^{-q_{T,b}a},\\
Q_{12}= &\left[-\frac{\alpha_{bl}}{\kappa_{bl}^+}\frac{r_{lb}^+}{\kappa_{lb}^+} \left(\Delta_{l}^{\text{II}} -e^{-2q_{T,l}t_l}      \right)e^{q_{T,l}t_l }
 +\frac{\alpha_{lb}}{\kappa_{lb}^+} \left(1+\Delta_{l}^{\text{III}}e^{-2q_{L,l}t_l} \right)e^{q_{L,l}t_l}\right] e^{-q_{T,b}a},\\
Q_{13}=&    \left(1 -\left[\Delta_{r}^{\text{II}}\right]^2e^{-2q_{T,r}t_r}\right)e^{ q_{T,r}t_r} +\frac{\alpha_{rb}\beta_{br}}
{r_{rb}^+r_{br}^+}\left(1-e^{-2q_{L,r}t_r}\right)e^{q_{L,r}t_r }, \\
Q_{14}=&  \frac{\alpha_{br}}{\kappa_{br}^+}\frac{r_{rb}^+}{\kappa_{rb}^+}  \left( 1-\Delta_{r}^{\text{II}} e^{-2q_{T,r}t_r}\right)e^{q_{T,r}t_r }+\frac{\alpha_{rb}}{\kappa_{rb}^+}\left(1+\Delta_{r}^{\text{III}}e^{-2q_{L,r}t_r}\right) e^{q_{L,r}t_r},  \\
Q_{21}=&   \left(1-\left[\Delta_{l}^{\text{II}}\right]^2e^{-2q_{T,l}t_l}  \right) e^{q_{T,l}t_l} +\frac{\alpha_{lb}}{r_{lb}^+}\frac{\beta_{bl}}{r_{bl}^+}\left(1-e^{-2q_{L,l}t_l} \right)e^{q_{L,l}t_l },   \\
Q_{22}=& \frac{ \alpha_{bl}}{\kappa_{bl}^+}\frac{r_{lb}^+}{\kappa_{lb}^+} \left(1-\Delta_{l}^{\text{II}} e^{-2q_{T,l}t_l}    \right)e^{q_{T,l}t_l }+\frac{\alpha_{lb} }{\kappa_{lb}^+}\left(1+\Delta_{l}^{\text{III}}e^{-2q_{L,l}t_l} \right) e^{q_{L,l}t_l},    \\
 Q_{23}=&\left[ -\Delta_r^{\text{II}}\left(1-e^{-2q_{T,r}t_r}\right) e^{q_{T,r}t_r} +\frac{\alpha_{rb}\beta_{br}}{r_{rb}^+r_{br}^+}\left(1-e^{-2q_{L,r}t_r}\right) e^{q_{L,r}t_r }\right]e^{-q_{T,b}a},\\
Q_{24}=& \left[-\frac{\alpha_{br}}{\kappa_{br}^+}\frac{r_{rb}^+}{\kappa_{rb}^+} \left(   \Delta_{r}^{\text{II}}- e^{-2q_{T,r}t_r}\right)e^{q_{T,r}t_r }+\frac{\alpha_{rb}}{\kappa_{rb}^+}\left(1+\Delta_{r}^{\text{III}}e^{-2q_{L,r}t_r}\right)e^{q_{L,r}t_r} \right] e^{-q_{T,b}a},\\
Q_{31}=&  \left[- \frac{\beta_{lb}}{r_{lb}^+}\left(1-\Delta_{l}^{\text{II}} e^{-2q_{T,l}t_l}    \right) e^{q_{T,l}t_l} -\frac{\beta_{bl}}{r_{bl}^+}\frac{\kappa_{lb}^+}{r_{lb}^+} \left(\Delta_{l}^{\text{III} } +e^{-2q_{L,l}t_l}  \right)e^{q_{L,l}t_l }\right] e^{-q_{L,b}a},\\
Q_{32}=&    \left[-\frac{ \alpha_{bl}\beta_{lb}}{\kappa_{bl}^+\kappa_{lb}^+}\left(1-e^{-2q_{T,l}t_l}\right)e^{q_{T,l}t_l }-\Delta_{l}^{\text{III}}\left( 1- e^{-2q_{L,l}t_l} \right) e^{q_{L,l}t_l}\right] e^{-q_{L,b}a},\\
 Q_{33}=&  \frac{\beta_{rb}}{r_{rb}^+}\left(1 -\Delta_{r}^{\text{II}} e^{-2q_{T,r}t_r}\right) e^{q_{T,r}t_r} +\frac{\beta_{br}}{r_{br}^+}\frac{\kappa_{rb}^+}{r_{rb}^+}\left(1+\Delta_{r}^{\text{III}}e^{-2q_{L,r}t_r}\right)e^{q_{L,r}t_r },  \\
Q_{34}=&  \frac{\beta_{rb}\alpha_{br}}{\kappa_{rb}^+\kappa_{br}^+}\left( 1- e^{-2q_{T,r}t_r} \right)e^{q_{T,r}t_r }+\left(1-\left[\Delta_{r}^{\text{III}}\right]^2e^{-2q_{L,r}t_r} \right)e^{q_{L,r}t_r},\\
Q_{41}=&    \frac{\beta_{lb}}{r_{lb}^+}\left(1-\Delta_{l}^{\text{II}} e^{ -2q_{T,l}t_l}    \right) e^{q_{T,l}t_l} +\frac{\beta_{bl}}{r_{bl}^+}\frac{\kappa_{lb}^+}{r_{lb}^+}\left(1+\Delta_{l}^{\text{III}} e^{-2q_{L,l}t_l}    \right)e^{q_{L,l}t_l },  \\
Q_{42}=&     \frac{\alpha_{bl}\beta_{lb}}{\kappa_{bl}^+\kappa_{lb}^+}\left(1-e^{-2q_{T,l}t_l}\right)e^{q_{T,l}t_l }+\left(1-\left[\Delta_{l}^{\text{III}}\right]^2e^{-2q_{L,l}t_l} \right)e^{q_{L,l}t_l},  \\
 Q_{43}=& \left[-\frac{\beta_{rb}}{r_{rb}^+}\left(1 -\Delta_{r}^{\text{II}} e^{-2q_{T,r}t_r}\right) e^{q_{T,r}t_r} -\frac{\beta_{br}}{r_{br}^+}\frac{\kappa_{rb}^+}{r_{rb}^+}\left(\Delta_{r}^{\text{III}}+e^{-2q_{L,r}t_r}\right)e^{q_{L,r}t_r }\right] e^{-q_{L,b}a},\\
Q_{44}=& \left[-\frac{\beta_{rb}\alpha_{br}}{\kappa_{rb}^+\kappa_{br}^+}\left( 1 - e^{-2q_{T,r}t_r} \right)e^{q_{T,r}t_r }-\Delta_{r}^{\text{III}} \left(1-e^{-2q_{L,r}t_r}\right) e^{q_{L,r}t_r}\right] e^{-q_{L,b}a};
\end{split}
\end{equation*}
and
$$Q_{\infty}=\begin{pmatrix} 0 & 0 & Q_{13} & Q_{14}\\
Q_{21} & Q_{22} & 0 & 0 \\
0 & 0 & Q_{33} & Q_{34}\\
Q_{41} & Q_{42} & 0 & 0 \end{pmatrix}$$is obtained by taking the limit $a\rightarrow \infty$ of $Q$.
In general, $\displaystyle \det\frac{Q}{Q_{\infty}}$ is a very complicated function of the distance $a$ between the plates. For the Casimir force acting between the plates, let $M_{ij}$ be the minor matrix of $Q$ obtained by deleting the $i^{\text{th}}$-row and $j^{\text{th}}$-column from $Q$. Then the TM contribution to the Casimir force acting between the plates is given by
\begin{equation*}\begin{split}
F_{\text{Cas}}^{\text{TM}} =-\frac{\pa E_{\text{Cas}}^{\text{TM}} }{\pa a}=\frac{\hbar}{4\pi^2 }\int_0^{\infty} \int_0^{\infty}
&\left\{\frac{q_{T,b}(Q_{11}M_{11}-Q_{12}M_{12}-Q_{23}M_{23}+Q_{24}M_{24})}{\det Q}\right.\\&\left.+\frac{q_{L,b}(Q_{31}M_{31}-Q_{32}M_{32}-Q_{43}M_{43}+Q_{44}M_{44})}{\det Q} \right\}k_{\perp}dk_{\perp}d\xi.
\end{split}\end{equation*}

In the massless $m\rightarrow 0$ limit, $q_{L,b}=q_{L,l}=q_{L,r}$. Therefore, $\alpha_{j_1j_2}= 0$ for all $j_1,j_2=b,l,r$, and $\Delta_{l}^{\text{III}}=\Delta_r^{\text{III}}=0$. The matrices $Q$ and $Q_{\infty}$ reduce  to   matrices of the form
\begin{align*}
Q^0 =\begin{pmatrix}Q_{11}^0 & 0&Q_{13}^0 & 0\\
Q_{21}^0 & 0&Q_{23}^0 & 0\\
Q_{31}^0 & 0&Q_{33}^0 & Q_{34}^0\\
Q_{41}^0 & Q_{42}^0 &Q_{43}^0 & 0\\
\end{pmatrix},\quad Q^0_{\infty}=\begin{pmatrix}0& 0&Q_{13}^0 & 0\\
Q_{21}^0 & 0&0 & 0\\
0& 0&Q_{33}^0 & Q_{34}^0\\
Q_{41}^0 & Q_{42}^0 &0 & 0\\
\end{pmatrix}.
\end{align*}Consequently,
\begin{equation}\label{eq7_22_8}\begin{split}
\det \frac{Q^0}{Q_{\infty}^0}=&-\frac{Q^0_{42}Q^0_{34}\left(Q^0_{11}Q^0_{23}-Q^0_{21}Q^0_{13}\right)}{
Q^0_{42}Q^0_{34}Q^0_{21}Q^0_{13}}=1-\frac{Q^0_{11}Q^0_{23}}{Q^0_{21}Q^0_{13}}
\\=& 1- \frac{\Delta_{l}^{\text{II}}\Delta_{r}^{\text{II}}\left(1-e^{-2q_{T,l}t_l}  \right)  \left(1-e^{-2q_{T,r}t_r}\right)}{
\left(1-\left[\Delta_{l}^{\text{II}}\right]^2e^{-2q_{T,l}t_l}  \right)\left(1 -\left[\Delta_{r}^{\text{II}}\right]^2e^{-2q_{T,r}t_r}\right)} e^{-2q_{T,b}a},
\end{split}\end{equation} where the massless limits of $\Delta_{j}^{\text{II}}, j=l,r,$ are
\begin{equation}\label{eq9_3_1}\lim_{m\rightarrow 0}\Delta_{j}^{\text{II}}=\frac{\mu_bq_{T,b}(\vep_j\mu_j\xi^2)-\mu_jq_{T,j}(\vep_b\mu_b\xi^2)}{\mu_bq_{T,b}(\vep_j\mu_j\xi^2)+\mu_jq_{T,j}(\vep_b\mu_b\xi^2)}
=\frac{\left.q_{T,b}(\xi,k_{\perp})\right|_{m=0}\vep_j(i\xi)-\left.q_{T,j}(\xi,k_{\perp})\right|_{m=0}\vep_b(i\xi)}
{\left.q_{T,b}(\xi,k_{\perp})\right|_{m=0}\vep_j(i\xi)+\left.q_{T,j}(\xi,k_{\perp})\right|_{m=0}\vep_b(i\xi)}.\end{equation} One finds that the massless limit of the  contribution to the Casimir energy from the superposition of the transverse modes of type II and longitudinal modes  is precisely the contribution from the TM modes in the massless case. This is another justification that we call the contribution from the superposition of the transverse modes of type II and longitudinal modes as TM   contribution. It is interesting to note that in the massless limit, the two polarizations for the massive field naturally reduce to only one polarization for the massless field.

\section{The Casimir effect on a pair of perfectly conducting  plates}\label{s6}

In this section,  we study the limits of the Casimir energy and Casimir force when the plates become perfectly conducting.  It is well-known that this can be obtained by the limit $\vep_l=\vep_r\rightarrow\infty$.   In this limit, we find that
$$q_{T,l},q_{T,r}\longrightarrow\infty;\quad  q_{T,b}= \sqrt{\vep_b(i\xi)\mu_b(i\xi)\xi^2+k_{\perp}^2+\frac{m^2c^2}{\hbar^2}},$$ $$ q_{L,b} = \sqrt{\frac{\xi^2}{c^2}+k_{\perp}^2+\frac{m^2 }{\vep_b(i\xi)\mu_b(i\xi)\hbar^2}},\quad q_{L,l}=q_{L,r}\longrightarrow \sqrt{\frac{\xi^2}{c^2}+k_{\perp}^2}:=q_0,$$$$ \Delta_{l}^{\text{I}}=\Delta_{r}^{\text{I}}\longrightarrow 1, \quad \Delta_{l}^{\text{II}}=\Delta_{r}^{\text{II}}\longrightarrow 1,\quad
\Delta_{l}^{\text{III}}=\Delta_{r}^{\text{III}}\longrightarrow \frac{q_0\left(q_{L,b}^2-k_{\perp}^2\right)-q_{L,b}\left(q_0^2-k_{\perp}^2\right)}
{q_{0}\left(q_{L,b}^2-k_{\perp}^2\right)+q_{L,b}\left(q_0^2-k_{\perp}^2\right)}:=\Delta,$$
$$\frac{\alpha_{lb}\beta_{lb}}{\kappa_{lb}^+r_{lb}^+}=\frac{\alpha_{bl}\beta_{bl}}{\kappa_{bl}^+r_{bl}^+}=
\frac{\alpha_{rb}\beta_{rb}}{\kappa_{rb}^+r_{rb}^+}=\frac{\alpha_{br}\beta_{br}}{\kappa_{br}^+r_{br}^+}
\longrightarrow \frac{k_{\perp}^2}{q_{T,b}}
\frac{q_0^2-q_{L,b}^2}{\left[q_{L,b}(q_0^2-k_{\perp}^2)+q_0(q_{L,b}^2-k_{\perp}^2)\right]}:=\Lambda.$$
For the TE contribution, it is easy to see from \eqref{eq7_22_1} that $\mathfrak{D}^{\text{TE}}\rightarrow 1$, and thus the  TE contribution to the  Casimir energy \eqref{eq7_22_2} reduces to
\begin{equation}\label{eq7_22_6}
\begin{split}
E_{\text{Cas}}^{\text{TE}}=&\frac{\hbar}{4\pi^2 }\int_0^{\infty}\int_0^{\infty}  \ln\left(1-e^{-2q_{T,b}(\xi,k_{\perp})a}\right)  k_{\perp}dk_{\perp}d\xi.
\end{split}
\end{equation} The TE contribution to the   Casimir force is then given by
\begin{equation}\label{eq7_26_2}
\begin{split}
F_{\text{Cas}}^{\text{TE}}=&-\frac{\hbar}{2\pi^2}\int_0^{\infty}\int_0^{\infty}  \frac{q_{T,b}(\xi,k_{\perp})}
{\exp\left(2q_{T,b}(\xi,k_{\perp})a\right)-1}k_{\perp}dk_{\perp}d\xi,
\end{split}
\end{equation}which is always attractive. It is interesting to note that in the perfect conductor limit, the TE contribution to the Casimir force is independent of the thicknesses of the plates.

The TM contribution is much more complicated.
With the help of computer, one can show that in the perfect conductor limit, the TM contribution to the Casimir energy \eqref{eq7_22_5} is given by
\begin{equation}\label{eq7_26_1}
\begin{split}
E_{\text{Cas}}^{\text{TM}} =& \frac{\hbar}{4\pi^2 }\int_0^{\infty} \int_0^{\infty} \ln \frac{W(\xi, k_{\perp})}{W_{\infty}(\xi, k_{\perp})} k_{\perp}dk_{\perp}d\xi,
\end{split}
\end{equation}where
$$W_{\infty}=\left([1-\Lambda]^2-[\Delta+\Lambda]^2e^{-2q_0t_l}\right)\left([1-\Lambda]^2-[\Delta+\Lambda]^2e^{-2q_0t_r}\right),$$and
\begin{equation}
\begin{split}
W=&W_{\infty}-\left([1-\Lambda^2]-[\Delta^2-\Lambda^2]e^{-2q_0t_l}\right)\left([1-\Lambda^2]-[\Delta^2-\Lambda^2]e^{-2q_0t_r}\right)e^{-2q_{T,b}a}\\
&-\left([1-\Lambda][\Delta-\Lambda]-[1+\Lambda][\Delta+\Lambda]e^{-2q_0t_l}\right)\left([1-\Lambda][\Delta-\Lambda]-[1+\Lambda][\Delta+\Lambda]e^{-2q_0t_r}\right)
e^{-2q_{L,b}a}\\
&+4\Lambda(1-\Delta)\left([1-\Lambda]+[\Delta+\Lambda]e^{-2q_0t_l}\right)\left([1-\Lambda]+[\Delta+\Lambda]e^{-2q_0t_r}\right)e^{-q_{T,b}a}e^{-q_{L,b}a}\\
&+(1-\Lambda)^2(\Delta+\Lambda)^2\left(1-e^{-2q_0t_l}\right)\left(1-e^{-2q_0t_r}\right)e^{-2q_{T,b}a}e^{-2q_{L,b}a}.
\end{split}
\end{equation}For an arbitrary background medium, one cannot split the Casimir energy \eqref{eq7_26_1} into contributions from two different polarizations. In the special case when the background material has refractive index $n_b=c\sqrt{\vep_b\mu_b}$ equal to one, e.g., if the perfectly conducting plates is placed in vacuum, then
$$q_{T,b}=q_{L,b}=\sqrt{\frac{\xi^2}{c^2}+k_{\perp}^2+\frac{m^2c^2}{\hbar^2}}:=q_m,$$ and $W/W_{\infty}$ simplifies drastically to
\begin{equation*}
\begin{split}
 \frac{W}{W_{\infty}}=&\left(1-e^{-2q_{m}a}\right)\left(1-\frac{(1-\Lambda)^2(\Delta+\Lambda)^2\left(1-e^{-2q_0t_l}\right)\left(1-e^{-2q_0t_r}\right)}
{\left([1-\Lambda]^2-[\Delta+\Lambda]^2e^{-2q_0t_l}\right)\left([1-\Lambda]^2-[\Delta+\Lambda]^2e^{-2q_0t_r}\right)}e^{-2q_ma}\right)\\=&\left(1-e^{-2q_{m}a}\right)
\left(1-\frac{D^2\left(1-e^{-2q_0t_l}\right)\left(1-e^{-2q_0t_r}\right)}{\left(1-D^2e^{-2q_0t_l}\right)\left(1-D^2e^{-2q_0t_r}\right)}e^{-2q_ma}\right),
\end{split}
\end{equation*}where
$$D=\frac{\Delta+\Lambda}{1-\Lambda}.$$ In this case, the TM contribution to the Casimir energy \eqref{eq7_26_1} can be written as a sum of two terms $E_{\text{Cas}}^{\text{TM, I}}$ and $E_{\text{Cas}}^{\text{TM, II}}$. $E_{\text{Cas}}^{\text{TM, I}}$ is the same as the TE contribution, obtained from \eqref{eq7_22_6} by setting $n_b\equiv 1$, and it corresponds to the massless TM contribution. The corresponding Casimir force is given by \eqref{eq7_26_2} with $n_b\equiv 1$, which is always attractive. The contribution to the Casimir energy  from the second TM polarization is given by
\begin{equation}\label{eq7_22_7}
\begin{split}
E_{\text{Cas}}^{\text{TM, II}} =& \frac{\hbar}{4\pi^2 }\int_0^{\infty} \int_0^{\infty} \ln\left\{1-\frac{D(\xi,k_{\perp})^2\left(1-e^{-2q_0t_l}\right)\left(1-e^{-2q_0t_r}\right)}
{\left(1-D(\xi,k_{\perp})^2e^{-2q_0t_l}\right)\left(1-D(\xi,k_{\perp})^2e^{-2q_0t_r}\right)}e^{-2q_ma}\right\} k_{\perp}dk_{\perp}d\xi.
\end{split}
\end{equation} Some simple computations give
\begin{equation*}
\begin{split}
D(\xi,k_{\perp})=\frac{\Delta+\Lambda}{1-\Lambda}=\frac{q_0\left(q_{m}^2-k_{\perp}^2\right)-q_{m}\left(q_0^2-k_{\perp}^2\right)+(q_0^2-q_m^2)\frac{k_{\perp}^2}{q_m}}
{q_{0}\left(q_{m}^2-k_{\perp}^2\right)+q_{m}\left(q_0^2-k_{\perp}^2\right)-(q_0^2-q_m^2)\frac{k_{\perp}^2}{q_m}}=\frac{q_m-q_0}{q_m+q_0}.
\end{split}\end{equation*}
The contribution to the Casimir force  from the second TM polarization is then given by
\begin{equation}\label{eq7_26_4}
\begin{split}
F_{\text{Cas}}^{\text{TM, II}} =& -\frac{\hbar}{2\pi^2 }\int_0^{\infty} \int_0^{\infty}q_m \left\{ \frac{\left(1-D(\xi,k_{\perp})^2e^{-2q_0t_l}\right)\left(1-D(\xi,k_{\perp})^2e^{-2q_0t_r}\right)} {D(\xi,k_{\perp})^2\left(1-e^{-2q_0t_l}\right)\left(1-e^{-2q_0t_r}\right)}
e^{2q_ma}-1\right\}^{-1} k_{\perp}dk_{\perp}d\xi.
\end{split}
\end{equation}Since $0<D<1$,
\begin{align*}
\frac{ 1-D^2e^{-2q_0t}}{D\left(1-e^{-2q_0t }\right)}-1=\frac{1-D^2e^{-2q_0t}-D\left(1-e^{-2q_0t }\right)}{D\left(1-e^{-2q_0t }\right)}
=\frac{(1-D)\left(1+De^{-2q_0t }\right)}{D\left(1-e^{-2q_0t }\right)}>0.
\end{align*}Therefore the sign of $F_{\text{Cas}}^{\text{TM, II}}$ is always negative, i.e., it is an attractive force.

Comparing \eqref{eq7_22_7} to \eqref{eq7_22_2}, we find that the second TM contribution \eqref{eq7_22_7} can be identified as    the TE contribution to the Casimir energy of a pair of dielectric plates due to a \emph{massless} electromagnetic field, where the permittivity of the dielectric plates is
$$\vep(\omega)=1-\frac{m^2c^4}{\hbar^2\omega^2}.$$This fact was observed in \cite{11}. In particular, in the limiting case of  perfectly conducting plates in vacuum, our general formula yields the result that was obtained by Barton and Dombey \cite{1}.

In order to gain more insight how the two TM polarizations split out in this special case, as was described in \cite{1}, let us return to the expression for the TM modes \eqref{eq7_22_9}.   In the case of perfectly conducting plates in vacuum, $p_{T,i}=p_{L,i}=p$ for $i=1,3,5$ and  $p_{T,2}, p_{T,4}\rightarrow \infty$. Let $p_{L,i}=p_0$ for $i=2,4$. Notice that $$ \frac{\omega^2}{c^2}=p^2+k_{\perp}^2+\frac{m^2c^2}{\hbar^2}=p_0^2+k_{\perp}^2.$$ The first boundary condition in \eqref{eq7_22_10} then implies that $\mathfrak{C}_2=\mathfrak{D}_2=\mathfrak{C}_4=\mathfrak{D}_4=0$. Since $\vep_2,\vep_4\rightarrow\infty$, \eqref{eq7_22_11} and \eqref{eq7_22_12} then imply that both the electric field and the magnetic field vanish in the plates. In the vacuum, $i=1,3,5$, the potentials are given by
\begin{equation}\label{eq7_23_1} \begin{split}
\left\{\begin{aligned}A_x=&\left[\left(\frac{k_{\perp}^2}{p } \mathfrak{C}_i+p  \mathfrak{E}_i\right)e^{ip x}-\left(\frac{k_{\perp}^2}{p } \mathfrak{D}_i+p \mathfrak{F}_i\right)e^{-ip x}\right] f_{\mathbf{k}_{\perp},\omega}(y,z,t)\\
A_y=&- k_2 \left[   \left(\mathfrak{C}_i-\mathfrak{E}_i\right)e^{ip x}+\left(\mathfrak{D}_i-\mathfrak{F}_i\right)e^{-ip x} \right]f_{\mathbf{k}_{\perp},\omega}(y,z,t)\\
A_z=&- k_3\left[  \left(\mathfrak{C}_i-\mathfrak{E}_i\right)e^{ip x}+\left(\mathfrak{D}_i-\mathfrak{F}_i\right)e^{-ip x} \right]f_{\mathbf{k}_{\perp},\omega}(y,z,t)\\
\phi=&\frac{c^2  }{\omega}(p^2+k_{\perp}^2)(\mathfrak{E}_ie^{ipx}+\mathfrak{F}_ie^{-ipx})f_{\mathbf{k}_{\perp},\omega}(y,z,t) \end{aligned}\right..
\end{split}\end{equation}In the plates, $i=2,4$, the potentials are given by
\begin{equation} \label{eq9_2_1}\begin{split}
\left\{\begin{aligned}A_x=& p_0 \left( \mathfrak{E}_i e^{ip_0 x}-  \mathfrak{F}_i e^{-ip_0 x}\right) f_{\mathbf{k}_{\perp},\omega}(y,z,t)\\
A_y=&  k_2 \left( \mathfrak{E}_i e^{ip_0 x}+  \mathfrak{F}_i e^{-ip_0 x}\right) f_{\mathbf{k}_{\perp},\omega}(y,z,t)\\
A_z=&  k_3 \left( \mathfrak{E}_i e^{ip_0 x}+  \mathfrak{F}_i e^{-ip_0 x}\right)f_{\mathbf{k}_{\perp},\omega}(y,z,t)\\
\phi=&\omega \left( \mathfrak{E}_i e^{ip_0 x}+  \mathfrak{F}_i e^{-ip_0 x}\right)f_{\mathbf{k}_{\perp},\omega}(y,z,t) \end{aligned}\right..
\end{split}\end{equation}
For $i=1,3,5$, renaming the variables by
\begin{equation*}\begin{split}
\mathfrak{C}_i'= \mathfrak{E}_i-\mathfrak{C}_i, \quad \mathfrak{E}_i'=\frac{k_{\perp}^2}{p}\mathfrak{C}_i+p\mathfrak{E}_i;\\
\mathfrak{D}_i'=\mathfrak{F}_i-\mathfrak{D}_i, \quad \mathfrak{F}_i'= \frac{k_{\perp}^2}{p}\mathfrak{D}_i+p\mathfrak{F}_i;
\end{split}\end{equation*}or equivalently,
\begin{equation*}
\begin{split}
\mathfrak{C}_i=-\frac{p^2}{p^2+k_{\perp}^2 }\mathfrak{C}_i'+\frac{p}{p^2+k_{\perp}^2 }\mathfrak{E}_i', \quad \mathfrak{E}_i=\frac{k_{\perp}^2}{p^2+k_{\perp}^2 }\mathfrak{C}_i'+ \frac{p}{p^2+k_{\perp}^2 }\mathfrak{E}_i';\\
\mathfrak{D}_i=-\frac{p^2}{p^2+k_{\perp}^2 }\mathfrak{D}_i'+\frac{p}{p^2+k_{\perp}^2 }\mathfrak{F}_i', \quad \mathfrak{F}_i= \frac{k_{\perp}^2}{p^2+k_{\perp}^2 }\mathfrak{D}_i'+\frac{p}{p^2+k_{\perp}^2 }\mathfrak{F}_i'.
\end{split}
\end{equation*}The formulas for the modes in the vacuum \eqref{eq7_23_1} then becomes
\begin{equation*} \begin{split}
\left\{\begin{aligned}A_x=& \left(  \mathfrak{E}_i'e^{ip x}- \mathfrak{F}_i'e^{-ip x}\right) f_{\mathbf{k}_{\perp},\omega}(y,z,t)\\
A_y=&  k_2    \left( \mathfrak{C}_i'e^{ip x}+\mathfrak{D}_i'e^{-ip x} \right)f_{\mathbf{k}_{\perp},\omega}(y,z,t)\\
A_z=&k_3 \left( \mathfrak{C}_i'e^{ip x}+\mathfrak{D}_i'e^{-ip x} \right)f_{\mathbf{k}_{\perp},\omega}(y,z,t)\\
\phi=&\frac{c^2  }{\omega} \left[\left(k_{\perp}^2\mathfrak{C}_i'+p\mathfrak{E}_i'\right)e^{ipx}+\left(k_{\perp}^2\mathfrak{D}_i'+p\mathfrak{F}_i'\right)e^{-ipx}\right]f_{\mathbf{k}_{\perp},\omega}(y,z,t) \end{aligned}\right.
\end{split}\end{equation*}for $i=1,3,5$; and   the boundary conditions \eqref{eq7_22_10} becomes
\begin{equation}\begin{split}
& p(\mathfrak{E}_j'e^{ipa_i}+\mathfrak{F}_j'e^{-ipa_i})+k_{\perp}^2 (\mathfrak{C}_j'e^{ipa_i}+\mathfrak{D}_j'e^{-ipa_i})
= \left(p_0^2+k_{\perp}^2\right)(\mathfrak{E}_{l}e^{ip_0a_i}+\mathfrak{F}_{l}e^{-ip_0a_i}),\\
& \mathfrak{E}_j'e^{ip a_i}- \mathfrak{F}_j'e^{-ip a_i}=p_0 \left( \mathfrak{E}_l e^{ip_0 a_i}-  \mathfrak{F}_l e^{-ip_0 a_i}\right),
\\
&\mathfrak{C}_j'e^{ip a_i}+\mathfrak{D}_j'e^{-ip a_i}= \mathfrak{E}_l e^{ip_0 a_i}+  \mathfrak{F}_l e^{-ip_0 a_i}.
\end{split}\end{equation}which follows from the continuities of $\phi$ and $A_x, A_y$. Here $l=2,4$. For $l=2$, $j=1$ and $i=1$ or $j=3$ and $i=2$. For $l=4$, $j=3$ and $i=3$ or $j=5$ and $i=4$. Substituting the third equation into the first equation, one obtains
$$p(\mathfrak{E}_j'e^{ipa_i}+\mathfrak{F}_j'e^{-ipa_i})=p_0^2(\mathfrak{E}_{l}e^{ip_0a_i}+\mathfrak{F}_{l}e^{-ip_0a_i}),$$which just reflects the continuity of $\displaystyle \frac{\pa A_x}{\pa x}$. The boundary conditions thus can be decomposed into two systems, one is the system
\begin{equation}\label{eq7_23_5}\begin{split}
&p(\mathfrak{E}_j'e^{ipa_i}+\mathfrak{F}_j'e^{-ipa_i})=p_0^2(\mathfrak{E}_{l}e^{ip_0a_i}+\mathfrak{F}_{l}e^{-ip_0a_i}),\\
& \mathfrak{E}_j'e^{ip a_i}- \mathfrak{F}_j'e^{-ip a_i}=p_0 \left( \mathfrak{E}_l e^{ip_0 a_i}-  \mathfrak{F}_l e^{-ip_0 a_i}\right)
\end{split}\end{equation}that reflects the continuities of $A_x$ and $\displaystyle \frac{\pa A_x}{\pa x}$, together with the conditions $\mathfrak{E}_1'e^{ipa_0}+\mathfrak{F}_1'e^{-ipa_0}=0$, $\mathfrak{E}_5'e^{ipa_5}+\mathfrak{F}_5'e^{-ipa_5}=0$ at the artificial boundaries $x=a_0$ and $x=a_5$; and the other one is the system
\begin{equation}\label{eq7_23_6}\mathfrak{C}_j'e^{ip a_i}+\mathfrak{D}_j'e^{-ip a_i}= \mathfrak{E}_l e^{ip_0 a_i}+  \mathfrak{F}_l e^{-ip_0 a_i}=\frac{p}{p_0^2}\left(\mathfrak{E}_j'e^{ipa_i}+\mathfrak{F}_j'e^{-ipa_i}\right) \end{equation}that reflect the continuities of $A_y, A_z$, together with the conditions $\mathfrak{C}_1'e^{ipa_0}+\mathfrak{D}_1'e^{-ipa_0}=0$, $\mathfrak{C}_5'e^{ipa_5}+\mathfrak{D}_5'e^{-ipa_5}=0$ at the artificial boundaries $x=a_0$ and $x=a_5$. In fact, \eqref{eq7_23_6} determines $(\mathfrak{C}_j',\mathfrak{D}_j'), j=1,3,5,$ from  $(\mathfrak{E}_l,\mathfrak{F}_l),l=2,4,$ but \emph{it does not determine uniquely}. One has the freedom to add any solutions satisfying
$$\mathfrak{C}_j'e^{ip a_i}+\mathfrak{D}_j'e^{-ip a_i}=0,\quad j=1,3,5,$$ corresponding to the trivial solution of $(\mathfrak{E}_j',\mathfrak{F}_j'), j=1,3,5,$  and $(\mathfrak{E}_l,\mathfrak{F}_l),l=2,4$. The nontrivial solutions of   $(\mathfrak{E}_j',\mathfrak{F}_j'), j=1,3,5,$  and $(\mathfrak{E}_l,\mathfrak{F}_l),l=2,4,$ satisfying \eqref{eq7_23_5} correspond to the type 3 continuum modes discussed in \cite{1}. For $j=1,3,5$, choosing $$\mathfrak{C}_j'=\frac{p}{p_0^2}\mathfrak{E}_j',\hspace{1cm} \mathfrak{D}_j'=\frac{p}{p_0^2}\mathfrak{F}_j',$$eq. \eqref{eq7_23_6} is satisfied, and we can fix the type 3 continuum modes to be
\begin{equation*} \begin{split}
\left\{\begin{aligned}A_x=& \left(  \mathfrak{E}_i'e^{ip x}- \mathfrak{F}_i'e^{-ip x}\right) f_{\mathbf{k}_{\perp},\omega}(y,z,t)\\
A_y=& \frac{p k_2}{p_0^2}    \left( \mathfrak{E}_i'e^{ip x}+\mathfrak{F}_i'e^{-ip x} \right)f_{\mathbf{k}_{\perp},\omega}(y,z,t)\\
A_z=&\frac{p k_3}{p_0^2} \left( \mathfrak{E}_i'e^{ip x}+\mathfrak{F}_i'e^{-ip x} \right)f_{\mathbf{k}_{\perp},\omega}(y,z,t)\\
\phi=&\frac{\omega p  }{p_0^2} \left( \mathfrak{E}_i' e^{ipx}+ \mathfrak{F}_i' e^{-ipx}\right)f_{\mathbf{k}_{\perp},\omega}(y,z,t) \end{aligned}\right.,
\end{split}\end{equation*}where $i=1,3,5$ for $a_{i-1}<x<a_i$. For $a_1<x<a_2$ and $a_3<x<a_4$, the modes are given by \eqref{eq9_2_1}. The boundary conditions satisfied by these modes are given by \eqref{eq7_23_5}. It is easy to verify that the contribution of these modes to the Casimir energy is given by $E_{ \text{Cas}}^{\text{TM, II}}$. For the trivial solution $\mathfrak{E}_j'=\mathfrak{F}_j'=0, j=1,3,5; \mathfrak{E}_l= \mathfrak{F}_l=0, l=2,4$, the potentials vanish identically on the plates. In the vacuum, we have three systems of equations:
\begin{equation*}
\begin{split}\left\{\begin{aligned}
&\mathfrak{C}_j'e^{ipa_{j-1}}+\mathfrak{D}_j'e^{-ipa_{j-1}}=0\\
&\mathfrak{C}_j'e^{ipa_{j}}+\mathfrak{D}_j'e^{-ipa_{j}}=0\end{aligned}\right.,\hspace{1cm} j=1,3,5.
\end{split}
\end{equation*}\begin{figure}
\epsfxsize=0.49\linewidth \epsffile{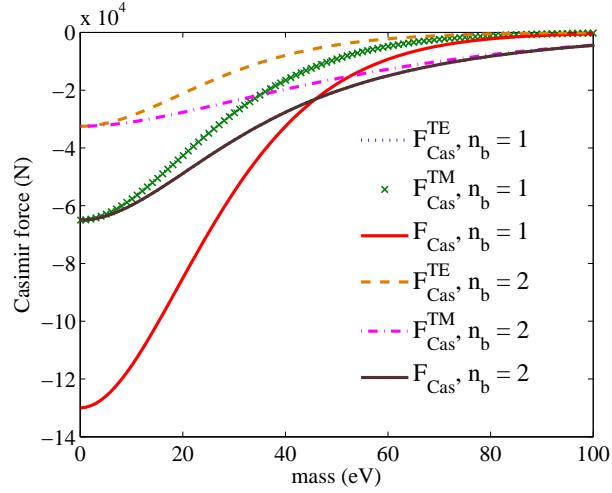}   \caption{\label{f3}The dependence of  the Casimir forces on the mass $m$ when the background medium has refractive index 1 and 2. }\end{figure}
These three systems are independent. In the limit $d_1=a_1-a_0$ and $d_5=a_5-a_4$ go to infinity, we only need to be concerned with the system with $j=3$, which has nontrivial solutions if and only if $\sin pa=0$, where $a=d_3=a_3-a_2$. In other words, $$p=\frac{\pi n}{a}, \quad n=1,2,\ldots,$$ and the corresponding potentials are given by
\begin{equation*}
\begin{split}
\left\{\begin{aligned}A_x=& 0\\
A_y=&  k_2   \sin \frac{\pi n (x-a_2)}{a_3-a_2}f_{\mathbf{k}_{\perp},\omega}(y,z,t)\\
A_z=&k_3\sin \frac{\pi n (x-a_2)}{a_3-a_2}f_{\mathbf{k}_{\perp},\omega}(y,z,t)\\
\phi=&\frac{c^2k_{\perp}^2  }{\omega} \sin \frac{\pi n (x-a_2)}{a_3-a_2}f_{\mathbf{k}_{\perp},\omega}(y,z,t) \end{aligned}\right.
\end{split}
\end{equation*} in the region $a_2<x<a_3$. This is precisely the type 2 discrete modes discussed in \cite{1}. Their contribution to the Casimir energy is $E_{ \text{Cas}}^{\text{TM, I}}$. Notice that they are totally different from the transverse modes of type II.

In FIG. \ref{f3}, we plot the Casimir forces  as a function of mass when the background medium has constant refractive index $n_b=1$ and $n_b=2$. In the  graph, we choose $a=t_l=t_r$=10nm.  From the graph, we can see that the TE contribution and the TM contribution to the Casimir force are approximately the same when the refractive index of the background medium is one. Since the first TM contribution to the Casimir force is equal to the TE contribution to the Casimir force in this case, we find that the second TM contribution to the Casimir force is negligibly small. However when the background medium has refractive index $n_b=2$, the graph shows that there is a significant difference between the TE and TM contributions. In fact, when $m$ gets larger, the total Casimir force is dominated by the TM contribution.

\section{The Casimir effect on a pair of infinitely permeable  plates}\label{s7}

In this section, we consider the limits of the Casimir energy and Casimir force when the plates become infinitely permeable, i.e.,  $\mu_l,\mu_r\rightarrow\infty$. In this
limit,
$$q_{T,l},q_{T,r}\longrightarrow\infty;\quad  q_{L,l}=q_{L,r}\longrightarrow \sqrt{\frac{\xi^2}{c^2}+k_{\perp}^2}:=q_0,$$$$ \Delta_{l}^{\text{I}}=\Delta_{r}^{\text{I}}\longrightarrow -1, \quad \Delta_{l}^{\text{II}}=\Delta_{r}^{\text{II}}\longrightarrow -1,\quad
\Delta_{l}^{\text{III}}=\Delta_{r}^{\text{III}}\longrightarrow \frac{q_0\left(q_{L,b}^2-k_{\perp}^2\right)-q_{L,b}\left(q_0^2-k_{\perp}^2\right)}
{q_{0}\left(q_{L,b}^2-k_{\perp}^2\right)+q_{L,b}\left(q_0^2-k_{\perp}^2\right)}:=\Delta,$$
$$\frac{\tilde{\alpha}_{lb}\tilde{\beta}_{lb}}{\tilde{\kappa}_{lb}^+\tilde{r}_{lb}^+}=\frac{\tilde{\alpha}_{bl}\tilde{\beta}_{bl}}{\tilde{\kappa}_{bl}^+\tilde{r}_{bl}^+}
=
\frac{\tilde{\alpha}_{rb}\tilde{\beta}_{rb}}{\tilde{\kappa}_{rb}^+\tilde{r}_{rb}^+}=\frac{\tilde{\alpha}_{br}\tilde{\beta}_{br}}{\tilde{\kappa}_{br}^+\tilde{r}_{br}^+}
\;\longrightarrow 0.$$From these, it is immediate to show that the TE contribution to the Casimir energy for infinitely permeable plates is given by \eqref{eq7_22_6}, the same as in the case of perfectly conducting plates.
It follows that the TE contribution to the Casimir force is given by \eqref{eq7_26_2}, which is always attractive.

For the TM contribution to the Casimir energy, one can show that
\begin{equation}
\begin{split}
E_{\text{Cas}}^{\text{TM}} =& \frac{\hbar}{4\pi^2 }\int_0^{\infty} \int_0^{\infty} \ln \mathcal{W}(\xi, k_{\perp}) k_{\perp}dk_{\perp}d\xi,
\end{split}
\end{equation}where
\begin{equation}
\label{eq9_2_2}\begin{split}\mathcal{W} =&\frac{(1-e^{-2q_{T,b}a})\left(\left[1- \Delta^2e^{-2q_0t_l}\right]\left[1-\Delta^2e^{-2q_0t_r}\right]-\Delta^2[1-e^{-2q_0t_l}][1-e^{-2q_0t_r}]e^{-2q_{L,b}a}\right)}{\left(1- \Delta^2e^{-2q_0t_l}\right)\left(1-\Delta^2e^{-2q_0t_r}\right)}\\=&(1-e^{-2q_{T,b}a})\left(1-\frac{\Delta^2(1-e^{-2q_0t_l})(1-e^{-2q_0t_r})}{\left(1- \Delta^2e^{-2q_0t_l}\right)\left(1-\Delta^2e^{-2q_0t_r}\right)}e^{-2q_{L,b}a}\right) .\end{split}\end{equation}
Therefore,
the TM contribution to the Casimir energy $E_{\text{Cas}}^{\text{TM}}$ can be split into the sum of two terms  $E_{\text{Cas}}^{\text{TM, I}}$ and $E_{\text{Cas}}^{\text{TM, II}}$. The first TM contribution to the Casimir energy $E_{\text{Cas}}^{\text{TM, I}}$ is the same as the TE contribution, given by \eqref{eq7_22_6}. Therefore the corresponding Casimir force is also always attractive. The second TM contribution to the Casimir energy and Casimir force are given respectively by
\begin{equation}\label{eq9_3_2}
\begin{split}
E_{\text{Cas}}^{\text{TM, II}} =& \frac{\hbar}{4\pi^2 }\int_0^{\infty} \int_0^{\infty} \ln \left(1-\frac{\Delta^2(1-e^{-2q_0t_l})(1-e^{-2q_0t_r})}{\left(1- \Delta^2e^{-2q_0t_l}\right)\left(1-\Delta^2e^{-2q_0t_r}\right)}e^{-2q_{L,b}a}\right) k_{\perp}dk_{\perp}d\xi,\\
F_{\text{Cas}}^{\text{TM, II}} =& -\frac{\hbar}{2\pi^2 }\int_0^{\infty} \int_0^{\infty}  q_{L,b} \left\{\frac{\left(1- \Delta^2e^{-2q_0t_l}\right)\left(1-\Delta^2e^{-2q_0t_r}\right)}{\Delta^2(1-e^{-2q_0t_l})(1-e^{-2q_0t_r})}e^{  2q_{L,b}a}-1\right\}^{-1} k_{\perp}dk_{\perp}d\xi.
\end{split}
\end{equation}When the refractive index of the background $n_b$ is a positive constant,  $q_{L,b}>q_0$. Therefore $0<\Delta<1$. It follows that the second TM contribution to the Casimir force $F_{\text{Cas}}^{\text{TM, II}} $ is also always attractive. This implies  that the Casimir force acting on a pair of infinitely permeable plates is always attractive.

Notice that
\begin{align*}
\Delta=\frac{q_0\left(q_{L,b}^2-k_{\perp}^2\right)-q_{L,b}\left(q_0^2-k_{\perp}^2\right)}
{q_{0}\left(q_{L,b}^2-k_{\perp}^2\right)+q_{L,b}\left(q_0^2-k_{\perp}^2\right)}=\frac{\sqrt{\frac{\xi^2}{c^2}+k_{\perp}^2}\hat{\vep}(i\xi)
-\sqrt{\hat{\vep}(i\xi)\frac{\xi^2}{c^2}+k_{\perp}^2}}{\sqrt{\frac{\xi^2}{c^2}+k_{\perp}^2}\hat{\vep}(i\xi)
+\sqrt{\hat{\vep}(i\xi)\frac{\xi^2}{c^2}+k_{\perp}^2}},
\end{align*}where $$\hat{\vep}(\omega)= 1-\frac{m^2c^2}{\vep_b(\omega)\mu_b(\omega)\hbar^2\omega^2}.$$
Comparing to \eqref{eq9_3_1}, we find that the second TM contribution to the Casimir energy \eqref{eq9_3_2}   can be identified with the TM contribution to the Casimir energy of a pair of dielectric plates with permittivity $\hat{\vep}$ that is due to a \emph{massless} vector field.

\begin{figure}[h]
\epsfxsize=0.5\linewidth \epsffile{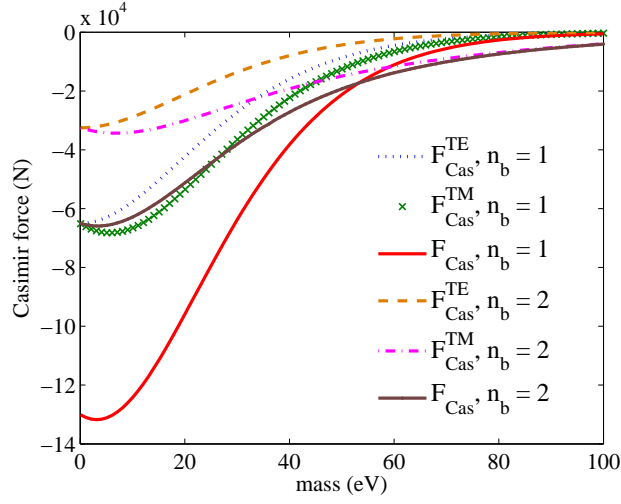}  \caption{\label{f4}The dependence of  the Casimir forces on the mass $m$ when the background medium has refractive index 1 and 2. }\end{figure}

In FIG. \ref{f4}, we plot the Casimir forces  as a function of mass when the background medium has constant refractive index $n_b=1$ and $n_b=2$. In this graph, we choose $a=t_l=t_r$=10nm.  We can see there is a significant difference between the TE and TM contributions to the Casimir force.

\section{The Casimir effect between a perfectly conducting plate and an infinitely permeable plate }\label{s8}

In this section, we consider the limits of the Casimir energy and Casimir force in   Boyer's setup \cite{23}, i.e.,    one plate is perfectly conducting and one plate is infinitely permeable. Without loss of generality, assume that the plate on the left is perfectly conducting, and the plate on the right is infinitely permeable,  i.e.,  $\vep_l,\mu_r\rightarrow\infty$. In this
limit,
$$q_{T,l},q_{T,r}\longrightarrow\infty;\quad  q_{L,l}=q_{L,r}\longrightarrow \sqrt{\frac{\xi^2}{c^2}+k_{\perp}^2}:=q_0,$$$$ \Delta_{l}^{\text{I}}\longrightarrow 1,\quad \Delta_{r}^{\text{I}}\longrightarrow -1, \quad \Delta_{l}^{\text{II}}\longrightarrow 1,\quad \Delta_{r}^{\text{II}}\longrightarrow -1,\quad
\Delta_{l}^{\text{III}}=\Delta_{r}^{\text{III}}\longrightarrow \frac{q_0\left(q_{L,b}^2-k_{\perp}^2\right)-q_{L,b}\left(q_0^2-k_{\perp}^2\right)}
{q_{0}\left(q_{L,b}^2-k_{\perp}^2\right)+q_{L,b}\left(q_0^2-k_{\perp}^2\right)}:=\Delta,$$
$$\frac{\tilde{\alpha}_{lb}\tilde{\beta}_{lb}}{\tilde{\kappa}_{lb}^+\tilde{r}_{lb}^+}=\frac{\tilde{\alpha}_{bl}\tilde{\beta}_{bl}}{\tilde{\kappa}_{bl}^+\tilde{r}_{bl}^+}
\longrightarrow \frac{k_{\perp}^2}{q_{T,b}}
\frac{q_0^2-q_{L,b}^2}{\left[q_{L,b}(q_0^2-k_{\perp}^2)+q_0(q_{L,b}^2-k_{\perp}^2)\right]}:=\Lambda,\quad
\frac{\tilde{\alpha}_{rb}\tilde{\beta}_{rb}}{\tilde{\kappa}_{rb}^+\tilde{r}_{rb}^+}=\frac{\tilde{\alpha}_{br}\tilde{\beta}_{br}}{\tilde{\kappa}_{br}^+\tilde{r}_{br}^+}
\;\longrightarrow 0.$$The TE contribution to the Casimir energy and Casimir force is then given respectively by
\begin{equation}\label{eq9_3_3}
\begin{split}
E_{\text{Cas}}^{\text{TE}}=&\frac{\hbar}{4\pi^2 }\int_0^{\infty}\int_0^{\infty}  \ln\left(1+e^{-2q_{T,b}(\xi,k_{\perp})a}\right)  k_{\perp}dk_{\perp}d\xi,
\\
F_{\text{Cas}}^{\text{TE}}=&\frac{\hbar}{2\pi^2}\int_0^{\infty}\int_0^{\infty}  \frac{q_{T,b}(\xi,k_{\perp})}
{\exp\left(2q_{T,b}(\xi,k_{\perp})a\right)+1}k_{\perp}dk_{\perp}d\xi.
\end{split}
\end{equation}It follows that the TE contribution to the Casimir force is always repulsive.

For the TM contribution, one can show that
\begin{equation}\label{eq9_3_4}
\begin{split}
E_{\text{Cas}}^{\text{TE}}=&\frac{\hbar}{4\pi^2 }\int_0^{\infty}\int_0^{\infty}  \ln  \frac{\mathbb{W}(\xi,k_{\perp})}{
\mathbb{W}_{\infty}(\xi,k_{\perp})}   k_{\perp}dk_{\perp}d\xi,
\end{split}
\end{equation}where
\begin{equation}\label{eq9_3_4}
\begin{split}
 \mathbb{W}_{\infty}  =&(1-\Delta^2e^{-2q_0t_l})\left([1-\Lambda]^2-[\Delta+\Lambda]^2e^{-2q_0t_r}\right)\\
 \mathbb{W}=&\mathbb{W}_{\infty}+(1-\Delta^2e^{-2q_0t_l})\left([1-\Lambda^2]-[\Delta^2-\Lambda^2]e^{-2q_0t_r}\right)e^{-2q_{T,b}a}
 -\Delta(1- e^{-2q_0t_l})\\&\times\left([1-\Lambda][\Delta-\Lambda]-[1+\Lambda][\Delta+\Lambda]e^{-2q_0t_r}\right)e^{-2q_{L,b}a}-\Delta(1-\Lambda)(\Delta+\Lambda)
 (1- e^{-2q_0t_l})(1- e^{-2q_0t_r})e^{-2q_{T,b}a}e^{-2q_{L,b}a}
\end{split}
\end{equation}
In this case, the TM contribution cannot be split into the sum of two   contributions even if we assume that the background medium have unity refractive index.

\begin{figure}[h]
\epsfxsize=0.5\linewidth \epsffile{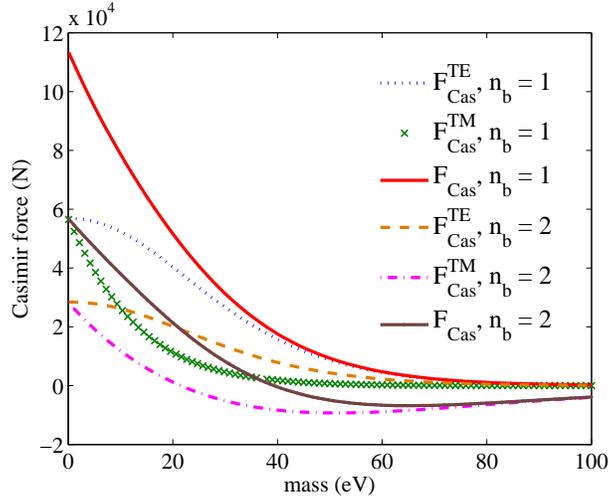}  \caption{\label{f5}The dependence of  the Casimir forces on the mass $m$ when the background medium has refractive index 1 and 2. }\end{figure}

In FIG. \ref{f5}, we plot the Casimir forces  as a function of mass when the background medium has constant refractive index $n_b=1$ and $n_b=2$. In this graph, we choose $a=t_l=t_r$=10nm.  From the graph, we find that the Casimir force is repulsive when the background medium has refractive index $n_b=1$. However, when the background medium has refractive index $n_b=2$, the Casimir force can change from repulsive to attractive when the mass increases. This shows that the change of mass can change the sign of the Casimir force.

\section{  Conclusion}\label{s9}
In this article, we have derived the Casimir energy and Casimir force acting on two parallel plates due to the vacuum fluctuations of a massive vector field. We assume that the two parallel  plates are made of real materials and they are placed in  a magnetodielectric background. With Maxwell's equations replaced by Proca equations, we argue that the correct boundary conditions for massive vector fields are the continuities of $\phi, \mathbf{A},   \pa_x A_x, \mathbf{H}_{\parallel}, \mathbf{E}_{\parallel}, \mathbf{B}_{\perp}$ and $\displaystyle \left(\pa_t\mathbf{D}-\frac{m^2c^2}{\mu\hbar^2}\mathbf{A}\right)_{\perp}$. The last one is equivalent to the continuity of $\mathbf{D}_{\perp}$ in the massless case. Not all the boundary conditions are independent. A set of independent boundary conditions is given by the continuities of $\mathbf{A}, \mathbf{H}_{\parallel}$ and the continuity of either $\phi$ or $\pa_x A_x$. The plane waves in an unbounded media can be divided into transverse waves of type I and type II and longitudinal waves, where the transverse waves of type I and type II are natural extensions of TE waves and TM waves in the massless case. For a system of several plane parallel layers of general magnetodielectric media, there are transverse modes of type I that satisfy all the boundary conditions. The contribution  to the Casimir energy from these modes is the natural generalization of the TE contribution in the massless case, and we also call it TE contribution. In general, there are no transverse modes of type II or longitudinal modes that satisfy all the boundary conditions. Therefore these two types of modes have to be combined.  The formula of their contribution  to the Casimir energy is very complicated. However, in the massless limit,  the two polarizations reduce to one polarization and the TM contribution to the Casimir energy in the massless case is reproduced. Therefore, we also call the contribution from the superposition of type II transverse modes and longitudinal modes the TM contribution.

In the limit where the plates become perfectly conducting, obtained by letting the permittivities of the plates tend to infinity, we write down explicitly the formulas for the TE and TM contributions to the Casimir energy. For general magnetodielectric background, the TM contribution cannot be split into two pieces. However, when the refractive index of the background is equal to unity, which happens for instance when the background is vacuum, the TM contribution can be split into two pieces, corresponding to the contribution from type 2 discrete modes and contribution from type 3 continuum modes discussed in the work of Barton and Dombey \cite{1}. In other words, our result  reproduces the result of Barton and Dombey \cite{1} for the special case they have considered. In this case, the Casimir force is always attractive.

In the limit where the plates become infinitely permeable, obtained by letting the permeabilities of the plates tend to infinity, the TM contribution to the Casimir energy can always be split into a sum of two terms. In this case, the Casimir force is always attractive when the background has constant refractive index.

To look for possible scenario that leads to repulsive force, we   consider the configuration proposed by Boyer \cite{23}, where one plate is perfectly conducting and one plate is infinitely permeable. In this case, we find that the TM contribution to the Casimir force cannot be split into a sum of two terms even when the background medium has unity refractive index. Numerical computation shows that the Casimir force can become attractive for nonzero masses when the refractive index of the background medium is not one. This shows that compare to the massless case, the Casimir effect of a massive vector field can behave very differently.

\end{document}